\documentclass{article}

\usepackage{amsmath,amssymb,amsthm,mathrsfs} 
\usepackage[pdftex,bookmarks,colorlinks,breaklinks]{hyperref}  

\title{Comment on \href{http://www.arxiv.org/abs/1205.3636}{``Can we measure structures to a precision better than the Planck length?''}, by Sabine Hossenfelder }
\author{S.~Doplicher\thanks{Dipartimento di Matematica, Universit\`a ``La Sapienza'', P.le A. Moro 5, 00185, Roma, Italy. E-mail: {\tt dopliche@mat.uniroma1.it.}}, G.~Piacitelli\thanks{SISSA, via Bonomea 265, 34136 Trieste, Italy. E-mail: {\tt gherardo@piacitelli.org}.}, L.~Tomassini\thanks{Dipartimento di Scienze, Universit\`a di Chieti-Pescara G. d'Annunzio, I-65127
. E-mail: {\tt tomassini@sci.unich.it}.}, S.~Viaggiu\thanks{Dipartimento di Matematica, Universit\`a ``Tor Vergata'', 
Via della Ricerca Scientifica 1, I-00133 Roma, Italy. E-mail: 
{\tt viaggiu@axp.mat.uniroma2.it}.}}

\begin{document}

\maketitle 
\begin{abstract}First principles do imply  a  non-zero minimal distance  
between events in spacetime, but no positive lower bound to the
precision  of the  measurement  of a {\itshape single coordinate}.
\end{abstract}

Part of the literature on Noncommutative Spacetime is based 
on a misunderstanding; namely, it takes as identical the following two distinct statements:
\begin{enumerate}
\item The localisation of a single space-time coordinate of an event cannot be 
    performed with precision higher than the Planck Length;

\item The Euclidean distance between two events in space-time cannot be smaller 
     than the Planck Length.
\end{enumerate}
The paper \cite{sabine} in question does not seem to clarify this confusion.

We believe, and stressed in several occasions, that the first statement 
is incorrect, while the second is correct; in a sense made precise in what follows.

Of course, nobody can say the final word based on a reliable theory of 
Quantum Gravity; however, we can investigate what presently known and 
accepted first principles allow, and what they do not.

We do not know any basic principle which characterises the concurrence of General Relativity 
and Quantum Mechanics, in a similar way as the locality principle does for Special Relativity 
and Quantum Mechanics~\cite{fate_of_locality}.

But the concurrence of Classical General Relativity and Quantum Mechanics points to a 
principle of stability of spacetime against localisation of events, 
formulated in~\cite{DFR}. 
There Space Time Uncertainty Relations (STUR) for the Minkowski space
(Cartesian) coordinates were deduced from that principle; 
those STUR suggested 
a model of Quantum Spacetime (QST) exactly implementing the STUR; this model is fully 
Poincar\'e covariant.

More precisely, the starting point was the following
\bigskip

\centerline{\bfseries Principle of Gravitational Stability against localisation of events:}

\begin{quote}
{\it The gravitational field generated by the concentration of energy  
required by the Heisenberg uncertainty principle to localise an event
in
spacetime should not be so strong to hide the event itself to any distant
observer - distant compared to the Planck scale.}
\end{quote}

This principle cannot prevent arbitrarily accurate measurements of {\it a single coordinate}, 
or even of two. For, if $a$ is the common value of two small but fixed space uncertainties, $b$ 
the value of the third, and we are clever enough to arrange that in our measurement the 
energy packed in that region, due to the Heisenberg principle, is uniformly spread, than a 
freshman computation shows that the associated Newton potential {\it tends to zero
} everywhere as $b$ 
grows to infinity and $a$ is kept fixed, no matter how small $a$ is.

The quantum gravitational corrections can hardly 
be relevant in that limit. 

The example of exact ``cosmic string'' solutions of the 
Einstein equations discussed in~\cite{sabine} is not strictly pertinent, as it postulates from the start, in 
the form of the metric, that the thickness of the string, and hence 
some space uncertainty, is exactly zero; but even if we took such a 
solution as an indication of what happens, in the example discussed 
above, where $a$ is (not zero, but) very small, if $a$ were kept fixed but 
$b$ would become very large the energy density per unit length of the string 
would cease to be Planckian and would actually {\it tend to zero}, and we would be brought 
back to the conclusion suggested by the Newtonian approximation, in agreement with our STUR.

Before continuing our discussion, some words are in order on what is meant
here by ``localising an event''.

In Classical General Relativity we deal with a four dimensional
manifold (spacetime) equipped with a Lorentzian metric. All points of the
spacetime can be labeled by four real numbers, namely their coordinates.

These coordinates describe the localisation of events. Clearly, this makes
sense even if the coordinates themselves are not observable quantities.

But even if they were classical observables, specialists of Quantum
Mechanics would like to see a quantum mechanical observable which measures
a single coordinate.        

Quantum Field Theorists, however, are well acquainted, since the work by
Bohr and Rosenfeld \cite{BR}, and especially thanks to the ideas of Rudolf
Haag \cite{haag}, with the idea that observables can only be derived from
fields, and fields at a point are meaningless.
Special combinations of field operators,
all smeared with test functions with support in the space--time region \(\mathscr O\),
describe ``local observables which can be measured within \(\mathscr O\)''.

If we could replace the region \(\mathscr O\) by a point, 
its coordinates would {\itshape not} be
observables, of course, but mere labels to specify a point in the
underlying classical manifolds. But measuring the corresponding local
observable, i.e. as above some combination of field operators,  would be a
way to localise some event in that point.

Bohr and Rosenfeld, as we recalled, taught us that this is not possible, and we know since years that
locality combined with {\itshape special} relativity forbids that, on precise
mathematical grounds, based on first principles.

The localisation region for an event, in other words, might be as tiny as
we like but with non zero sizes.

That is, as long as General Relativity is {\itshape not} 
taken in account, i.e.\ we disregard
gravitational forces between elementary particles as exceedingly weak, we
can still talk of localising events in {\itshape arbitrarily small} 
regions \(\mathscr O\), by
means of measurements of local observables, along the above lines. This
gives a meaning to ``points'' in a classical manifold as idealised
localisation of events.

These very well known and very elementary considerations allow us to talk
of Minkowski space (or, in presence of an {\itshape external} background
gravitational field,  of an Einstein manifold), as a {\itshape classical
background} where Quantum Field Theory is formulated.

We must repeat: the points of such a manifold are {\itshape not} 
described by some
universal Quantum Field Theoretic observables called ``coordinates'', 
but they are
accessible to observation, with arbitrarily high precision, as attributes
of local observables (localisation), in the above sense. This will be more
and more accurate the better the state we produce with such a measurement
is localised in the desired region. Thus the Quantum Field Theoretic concept of
{\itshape localised state} \cite{haag} is crucial.

For instance, the state describing the result of a measurement localizing
an event around some point
\(x\) in
Minkowski space with uncertainties in the coordinates  \(\Delta x_\mu, \mu
= 0, 1, 2, 3\), will be represented by a vector obtained by the action on
the vacuum of some appropriate combinations of  field operators, smeared
out with test functions with support---say---in a region of rectangular
shape in Minkowski space, centered around the point \(x\), with sides of
size \(\Delta x_\mu\).

With all this in mind, the core of our message is:

\begin{quote}
If the gravitational forces are {\itshape not} 
disregarded, the localisation of an
event, in the sense described here above, cannot be realised unless the
uncertainties in the coordinates of the event, described as in the above
example, obey some specific uncertainty relations, derived from the
concurrence of Quantum Mechanics and Classical General Relativity.
\end{quote}

These STUR lead to the Planck length as a lower bound to the Euclidean
distance between two events, but  the uncertainty of a single
coordinate does not have a priori a nonzero lower bound.

This strongly suggests to abandon the concept of a classical Einstein
Manifold as a geometric background, in favour of a non commutative
manifold, describing Quantum Spacetime. So far this procedure has been
followed through only in the case of Quantum Minkowski Space \cite{DFR,ultraviolet,4volume}.

But in general, we might expect that the form of Noncommutativity should
not be a priori assigned, but part of the dynamics \cite{texture}.


How do our comments here compare to the ``generalised 
uncertainty principle'', is a point which also has been often emphasised 
(see e.g.~\cite{texture}; or~\cite{D}), where you can read:
\begin{quote}
``There is no a priori lower limit on the precision in the measurement
of any single coordinate (it is worthwhile to stress once more that the
apparently opposite conclusions, still often reported in the literature,
are drawn under the implicit assumption that all the space coordinates of
the event are simultaneously sharply measured);''
\end{quote}
or, more recently,~\cite{4volume}, where you can read, about the same fact:
\begin{quote}
``This does not conflict with the famous Amati Ciafaloni Veneziano
Generalized Uncertainty Relation: all the derivations we are aware of (see
e.g.[\ldots] ) implicitly assume that all space coordinates of the event are
measured with uncertainties of the same order of magnitude; in which case
they agree with our Spacetime Uncertainty Relations.''
\end{quote}
For a more detailed discussion, see the review article~\cite{GP}.

That implicit hypothesis seems to have been noticed in~\cite{sabine}, but the confusion minimal 
uncertainty in a coordinate / minimal length / minimal volume seems to persist.

Note that the STUR proposed in~\cite{DFR} \footnote{Strangely enough, the 
second relation is totally ignored in \cite{sabine}}, namely, in absolute units,

\begin{equation}
\label{dopleq12}
\Delta q_0 \cdot \sum \limits_{j = 1}^3 \Delta q_j \gtrsim 1 ; \sum
\limits_{1 \leq j < k \leq 3 } \Delta q_j  \Delta q_k \gtrsim 1 .
\end{equation}  

of course imply that each $\Delta q_j$ is greater than one (i.e. of the Planck length) 
when they all coincide, for  $j = 1,2,3$.  

Thus, the argument of 
Bronstein, and, if properly interpreted in view of the implicit 
hypothesis that all space uncertainties agree, the results of Mead, 
Amati--Ciafaloni--Veneziano, (referred to in~\cite{sabine,4volume,GP}) 
are not at all contradicted by these STUR.

Moreover these STUR do lead to assertion (ii) (though assertion (i) is 
negated).

To make this point precise, note first that if one sticks to the 
uncertainty relations, no sharp assertion on the geometric operators, 
as distance, area, 3- or 4-volume, is possible; to that effect an 
explicit mathematical model of spacetime must be assumed (or derived 
from dynamics) where the corresponding analytical expressions 
(operators, in models of Quantum Spacetime) can be defined and 
analysed.

Within the ``basic model'' of QST, which exactly implements the above 
STUR, one can exactly analyse the operators describing distance 
between two independent events~\cite{ultraviolet}, or, considering up to five independent 
events, area, three volume or four volume operators~\cite{4volume}.

The result is that the sum of the square moduli of all components, for   {\it each} of those operators,  (the square 
Euclidean distance,\ldots, the square modulus of the pseudoscalar four volume operator) is
bounded below by some positive constant of order 1 in Planck units - {\it in all Lorentz frames } 
~\cite{4volume} (although the model is fully Poincar\'e covariant; problems with Lorentz covariance arise only 
if dealing with {\it interactions } between quantum fields on QST~\cite{DFR, ultraviolet}.

Note that obvious modifications  of the computations in \cite[Section 3]{DFR}
lead to a similar lower bound (a factor \(1/2\) smaller, i.e. exactly
\(2\)
in
Planck units) for the square Euclidean spatial distance, i.e. for the sum of
the
square moduli of the differences of the space coordinates of the two
events.

The result on distances tells precisely that you cannot ``measure 
structures to a precision better that the Planck length''.

Are we asserting that we {\it can } measure a given coordinate of an event as 
precisely as we wish?

Of course not: we are asserting that no presently known argument or 
principle seems to forbid it.

The claim, repeatedly asserted, that we do
not know whether there is a positive lower bound on the accuracy with which a single
coordinate can be measured, apparently received very little attention; we would like to 
acknowledge that the paper~\cite{sabine} is  an exception.

Note that in all the quoted arguments it was important to talk of 
localisation of {\it events}, not of {\it particles}: by Heisenberg's principle, the 
uncontrolled amount of energy transmitted to the system by such an 
accurate measurement would create uncontrollable amounts of photons and 
particle-antiparticle pairs.

The well known fact we just mentioned takes a precise form in Quantum Field Theory, in the light of the principle of locality. 

Typically, localised states are obtained from the vacuum by acting with an isometric operator, which lies in a local algebra of fields. Coherent states, with components with arbitrarily many particles, are examples. But  single particle states are never of that form.\footnote  {Even if they are  in the Newton-Wigner or in the Foldy Wouthuysen class, their localisation will be merely the expression of a geometric relation between them, and not of the relativistic localisation: the expectation value of a local observable in such a state will never have the characteristic property, to become exactly the vacuum expectation value, when the observable is translated away  into the spacelike complement of some localisation region. It will merely converge to it, for large spacelike translations, with a characteristic distance of order of the Compton wave length of the particle.}

If one misses these points, one feeds confusions, as that between 
distance resolution and uncertainty in a single coordinate.

The limitation 2. was deduced in~\cite{DFR, ultraviolet, 4volume} from the Basic Model of QST based on the 
STUR; the weakness of the linear approximation (common to so many other results; mitigated, however, by the consistency with exact solutions as Schwarzschild's or Kerr's), made to derive the STUR in the 
starting point of~\cite{DFR}, was partly resolved in~\cite{TV}, based on the Hoop conjecture, and, with 
the limitation to spherical symmetry, but without use of the the Hoop conjecture, in~\cite{DMP}.

Similarly, we are {\it not } asserting that the Euclidean distance 
between two 
events in spacetime {\it can} be as small as the Planck length; we are only 
saying that, under the specified hypothesis, it cannot be smaller.

\end{document}